# Quantum efficiency enhancement of bialkali photocathodes by an atomically thin layer on substrates

*Hisato Yamaguchi\*, Fangze Liu, Jeffrey DeFazio, Mengjia Gaowei, Lei Guo, Anna Alexander, Seong In Yoon, Chohee Hyun, Matthew Critchley, John Sinsheimer, Vitaly Pavlenko, Derek Strom, Kevin L. Jensen, Daniel Finkenstadt, Hyeon Suk Shin, Masahiro Yamamoto, John Smedley, Nathan A. Moody*

Dr. H.Y., Dr. F.L., Dr. A.A., Dr. V.P., Dr. J.Smedley, Dr. N.A.M.
Los Alamos National Laboratory, P.O. Box 1663, Los Alamos, New Mexico 87545, U.S.A.
E-mail: hyamaguchi@lanl.gov
Dr. J.D.
Photonis Defense Inc., 1000 New Holland Ave., Lancaster, PA 17601, U.S.A.
Dr. M.G., Dr. J.S.
Brookhaven National Laboratory, P.O. Box 5000, Upton, New York 11973, U.S.A.
Dr. L.G.
Nagoya University, Furo, Chikusa, Nagoya, 464-8601, Japan
Mr. S.I.Y., Ms. C.H., Prof. Dr. H.S.S.
Department of Chemistry and Department of Energy Engineering, Ulsan National Institute of Science and Technology, Ulsan 44919, Republic of Korea
Mr. M.C., Prof. Dr. D.F.
U.S. Naval Academy, Stop 9c, 572c Holloway Rd., Annapolis, MD 21402, U.S.A.
Dr. D.S.
Max Planck Institute for Physics, Föhringer Ring 6, D-80805 Munich, Germany
Dr. K.L.J.
Naval Research Laboratory, 4555 Overlook Ave. SW, Washington DC 20375, U.S.A.
Dr. M.Y.
High Energy Accelerator Research Organization (KEK), 1-1 Oho, Tsukuba, Ibaraki 305-0801, Japan

**Abstract**

We report quantum efficiency (QE) enhancements in accelerator technology relevant antimonide photocathodes ($K_2CsSb$) by interfacing them with atomically thin two-dimensional (2D) crystal layers. The enhancement occurs in a reflection mode, when a 2D crystal is placed in between the photocathodes and optically reflective substrates. Specifically, the peak QE at 405 nm (3.1 eV) increases by a relative 10 %, while the long wavelength response at 633 nm (2.0 eV) increases by a relative 36 % on average and up to 80 % at localized "hot spot" regions when photocathodes are deposited onto graphene coated stainless steel. There is a similar effect for photocathodes deposited on hexagonal boron nitride monolayer coatings using nickel substrates. The enhancement does not occur when reflective substrates are replaced with optically transparent sapphire. Optical transmission, X-ray diffraction (XRD) and X-ray fluorescence (XRF) revealed that thickness, crystal orientation, quality and elemental stoichiometry of photocathodes do not appreciably change due to 2D crystal coatings. These results suggest optical interactions are responsible for the QE enhancements when 2D crystal sublayers are present on reflective substrates, and provide a pathway toward a simple method of QE enhancement in semiconductor photocathodes by an atomically thin 2D crystal on substrates.

# 1. Introduction

Besides the well-known application of imaging bones in human bodies, X-rays enable powerful and pervasive tools with countless uses throughout science and technology. They are used to answer many important materials related problems in our society, ranging from development of new medicines for curing cancers to development of high performance batteries for automobile industries, and the development of lightweight and high mechanical strength materials for space missions. Their sub-nanometer wavelengths allow structural information of materials to be probed at atomistic precisions, providing unique capabilities inaccessible to other regions of the electromagnetic spectrum.

The only instruments that are currently capable of generating the high brightness and coherent X-ray beams required for atomic scale material investigations are electron accelerator facilities. An emergent problem however is that the performance requirements on the scientific frontier of these investigations dramatically outstrip the capabilities of present state-of-the-art electron sources and cathode technologies[1-3]. The high demand for increasingly high performance electron beams is such that U.S. department of energy (DOE) commissioned studies have repeatedly identified electron sources as a critical risk area, forming one of the highest accelerator R&D priorities for the next decade[1-4].

There are multiple approaches to this problem. One challenging avenue is to shield high quantum efficiency (QE) bialkali antimonide photocathodes with graphene to meet the required transformational advances in lifetime and efficiency simultaneously[5, 6]. This type of approach is important in meeting an ultimate goal, however, the process could take many years and the outcome is not necessarily guaranteed. Therefore, pursuing alternate approaches in parallel is beneficial in mitigating the development risks. One such focus is surface passivation and/or the engineering of substrates for semiconductor photocathodes. For example, it is known that surface treatments for metallic substrates play an important role in securing maximum QE of deposited photocathodes. Only limited literature is available on this topic[7-9], but the consensus indicates benefits from electrochemically etching the surfaces to smoothen morphology (i.e. electropolishing) and further chemical passivation by exposing them to gases at elevated temperatures, etc. More recent approaches in this direction include use of semiconductor single crystals such as silicon and gallium arsenide for clean and atomically controlled surfaces[10-12], the use of different metal surfaces for reflectivity enhancement[13], and implementation of nanostructures[14, 15].

Here, we demonstrate that atomically thin two-dimensional (2D) crystals such as graphene and hexagonal boron nitride monolayers can play both roles of surface passivation and engineering of metal substrates. The generalized hypothesis driving this approach was as follows. 1) The atomic thinness of a graphene or hexagonal boron nitride monolayer is suited to coat any type of substrate surfaces due to its mechanical flexibility[16-20], 2) they can withstand the typical substrate cleaning processes of ~500 °C due to its thermal stability[21-24], 3) they provide chemically inert surfaces for possible highly-crystalline photocathode growth based on their dangling bond-free atomic structures[25-29], and 4) they are optically transparent in visible region (only 2.3 % absorption per monolayer for graphene) thus any QE decrease due to their light absorption should be minimal.

A particular photocathode material of interest is potassium cesium antimonide ($K_2CsSb$) because it possesses one of the highest visible QEs with a peak that can exceed 20 % at 3 eV, yet does not require the extremely deep operating vacuum of ~$10^{-11}$ Torr as that of activated gallium arsenide (GaAs: Cs-O)[30]. Specific results that we achieved include an unexpected QE enhancement compared to a non-coated case when 2D crystals are coated on metallic substrates. The enhancement was a relative 10 % on average at 405 nm (3.1 eV) and a relative 36 % at 633 nm (2.0 eV), with up to 80 % at localized "hot" spots. Our proposed structure of coating vanishingly thin layers on reflective substrates serves as a milestone towards a novel method to enhance the QE of accelerator-relevant semiconductor photocathodes.

## 2. Results and Discussion
### 2.1 Quantum efficiency maps of bialkali antimonide photocathodes on atomically thin 2D crystal coated substrates

*2.1.1 Graphene synthesis, hexagonal boron nitride monolayer and phototube fabrications*

The graphene and hexagonal boron nitride monolayer used in this study was grown by chemical vapor deposition (CVD). We synthesized graphene and confirmed its thickness to be monolayer with minimal structural defects (**Figure 1** (a)). Specifically, Raman spectroscopy showed a 2D/G peak ratio of ~3, where a 2D/G value of higher than ~2 is accepted as an indication of a monolayer[27]. There was no observable D peak at ~1350 cm$^{-1}$ indicating the structural defect induced vibration mode in graphitic materials. The graphene was further characterized by atomic force microscopy (AFM), which showed continuous films with monolayer thickness of ~0.5 nm. We purchased hexagonal boron nitride monolayer (see Experimental Section for details). For one of the phototubes, graphene film was transferred onto stainless steel and hexagonal boron nitride film were transferred onto nickel mesh grids (Figure 1 (b)). For another phototube, both graphene and hexagonal boron nitride monolayer films were transferred onto sapphire substrate (Figure 1 (c)). All of the transfers were performed using an established polymer-supported wet-based method. After removal of the polymer-support in acetone baths and drying, each set of 2D crystal-coated substrate was installed into vacuum tube assemblies. Potassium cesium antimonide (K$_2$CsSb) photocathodes were then deposited on the films and permanently sealed (Figure 1 (b), (c)). Sapphire windows with patterned metal grids were used as anodes to collect photoelectrons. The QE of the photocathodes was measured in reflection mode; i.e. illuminating from photocathode side and collecting emitted electrons from the same side (Figure 1 (d)). The vacuum phototube allows a unique opportunity for long-term stability that is inaccessible in dynamic pumping environments. The design used here also allows for routine and repeatable QE and optical measurements of the photocathodes of interest.

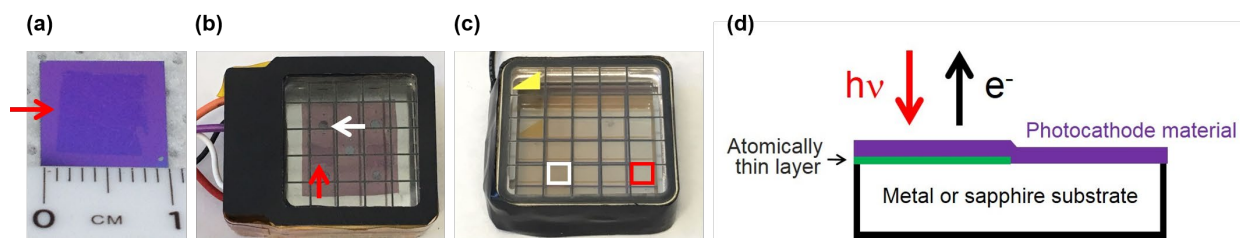

**Figure 1.** (a) Photograph of graphene transferred onto SiO$_2$/Si substrate. Red arrow indicates an edge of graphene film. (b) Photograph of K$_2$CsSb photocathode deposited on graphene coated stainless steel substrate and hexagonal boron nitride coated nickel mesh grid. Red and white arrows indicate regions of graphene and hexagonal boron nitride coating, respectively. (c) Photograph of K$_2$CsSb photocathode deposited on graphene and hexagonal boron nitride coated sapphire substrate. Red and white squares indicate regions of graphene and hexagonal boron

nitride coating, respectively. (d) Side view schematic of photocathode structure and photoemission measurements performed in this study.

*2.1.2 Quantum efficiency maps of photocathodes on graphene coated reflective substrates*

**Figure 2** (a) is a top view schematic of a photocathode used in this study. Specifically, we prepared a chemically passivated stainless steel substrate with a monolayer graphene region and grew a $K_2CsSb$ photocathode on it using a conventional sequential deposition. Figure 2 (b) is a QE map obtained by rastering a 405 nm light emitting diode (LED) with spot size of ~0.2 mm. An overall QE well in excess of 15 % was achieved over the 4 mm x 4 mm sample area, which indicates that our $K_2CsSb$ photocathodes are of high quality. What is immediately evident is an enhanced QE in the region with graphene coating. This region has a mean QE of ~ 20 % (standard deviation: 0.53 %) in contrast to ~18 % (standard deviation: 0.66 %) at the region without coating (Figure 2 (c)), or a relative 10 % increase compared to the uncoated surface. To resolve more detailed features, we performed high spatial resolution QE mapping using a focused laser of 350 nm spot size, using the same photon energy of the LED at 405 nm. Fine features at the interface that resemble optical microscopy images of the graphene coating are observed (Figure 2 (d)), supporting the notion that the graphene monolayer is responsible for the QE enhancement. The average QE difference between the regions with and without graphene was a relative ~ 8 %, consistent with the measurements in Figure 2 (b).

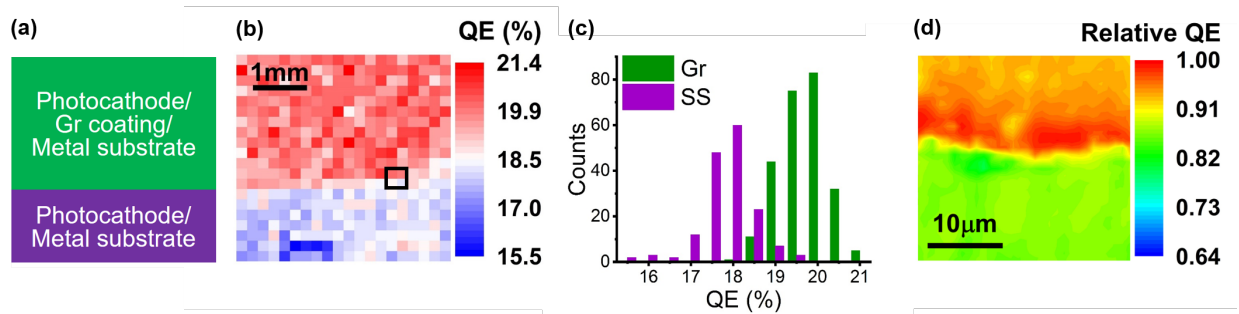

**Figure 2.** (a) Top view schematic of our photocathode structure. (b) 405 nm illuminated QE map of $K_2CsSb$ photocathodes with 0.2 mm spatial resolution. (c) Statistics of QE in (b) by pixel counts. Gr and SS labels represent pixels in regions with and without graphene coating in (b), respectively. (d) Enlarged region of the black square in (b) with intensity normalized to the maximum value.

*2.1.3 Quantum efficiency maps of photocathodes on graphene coated transparent substrates*

A role of substrate in the observed QE enhancement can be studied by making a comparison on optically transparent substrates. To do so, we fabricated a separate vacuum phototube with a sapphire substrate for the photocathode. The sapphire substrate similarly contained a region of monolayer graphene coating. **Figure 3** (a) is the top view schematic, which indicates our graphene coating. We obtained a QE map by the same configuration as previously described in the reflective substrate case, which is to raster a 405 nm light emitting diode with spot size of ~0.2 mm over the 4 mm x 4 mm sample area (Figure 3 (b)). An overall QE of >15 % was achieved similar to a reflective substrate case, indicating again that our $K_2CsSb$ photocathodes are of high quality. In sharp contrast to the stainless steel substrate, however, we observed an opposite effect of coating on the QE, i.e. the QE *decreased* for the region with graphene coating in this case where the substrate is optically transparent. The decrease was relative ~13 % on average with mean QE of ~16 % (standard deviation: 1.1 %) and ~18 % (standard deviation: 1.2 %) for the regions with and without the coating, respectively (Figure 3 (c)).

The result strongly suggests that an origin of enhanced QE is due to optical interactions between the coating, reflective substrate, and $K_2CsSb$ photocathodes. If the QE enhancement was due to other factors such as improvement of $K_2CsSb$ photocathodes crystal quality, then QE enhancement by the coating should occur regardless of substrates being reflective or optically transparent. We confirmed that the photocathode thicknesses do not change due to the coating (Figure 4 (c)).

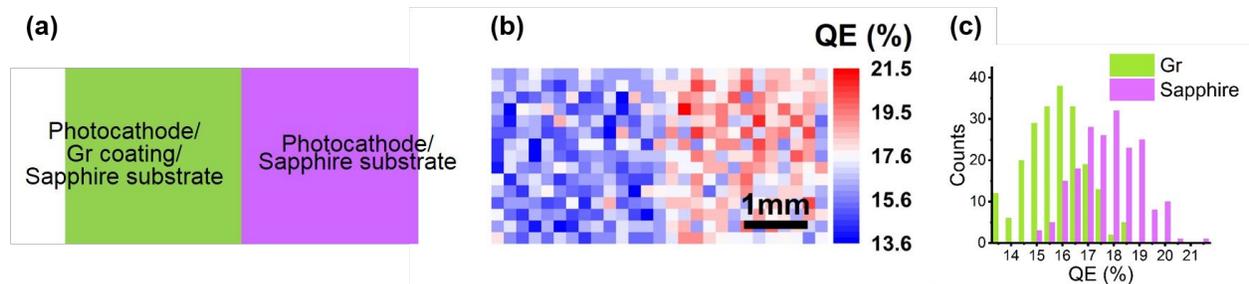

**Figure 3.** (a) Top view schematic of our photocathode structure. (b) Corresponding QE map taken by 405 nm illumination with 0.2 mm spatial resolution. Scale bar is 1 mm. (c) Statistics of QE in (b) by pixel counts. Gr and Sapphire labels represent pixels in regions with and without graphene coating in (b), respectively.

*2.1.4 Quantum efficiency maps of photocathodes on hexagonal boron nitride monolayer coated reflective substrates*

Our next question was whether the QE enhancement with the 2D crystal sublayer was specific to graphene. If the QE enhancement can be achieved for other compositions of 2D crystals, then our findings lead to a proposal of novel photocathode structure that depends less on a material type. To this end, we studied a case for CVD hexagonal boron nitride (hBN) monolayers, which were synthesized in similar conditions as that of graphene. Specifically, we transferred three individual monolayers of CVD grown hBN onto nickel substrate, then deposited $K_2CsSb$ photocathode on it in the same process as previously described for the stainless steel/graphene films. **Figure 4** (a) shows a high spatial resolution QE map of the $K_2CsSb$ photocathode on nickel substrate taken using focused laser with wavelength of 405 nm and spatial resolution of 350 nm. In comparison, Figure 4 (b) is the reference QE map of the simultaneously deposited $K_2CsSb$ photocathode on stainless steel without hBN. A mean QE nearing 20 % (standard deviation: 1.0 %) was achieved for the photocathode with hBN sublayer whereas it was less than 18 % (standard deviation: 0.31 %) for the bare stainless steel counterpart. This QE enhancement factor is in the similar range as that of graphene case, demonstrating that the QE enhancement can be achieved by atomically thin coatings other than graphene, and thus suggesting the possibility of a generalized method for enhancing the QE. When we checked optical transmission spectra of $K_2CsSb$ photocathodes on bare, graphene coated, and hexagonal boron nitride coated sapphire substrates (Figure 4 (c)), we only saw negligible difference between them that matches our model calculations, which indicates that $K_2CsSb$ photocathode thickness does not change due to 2D crystal coatings.

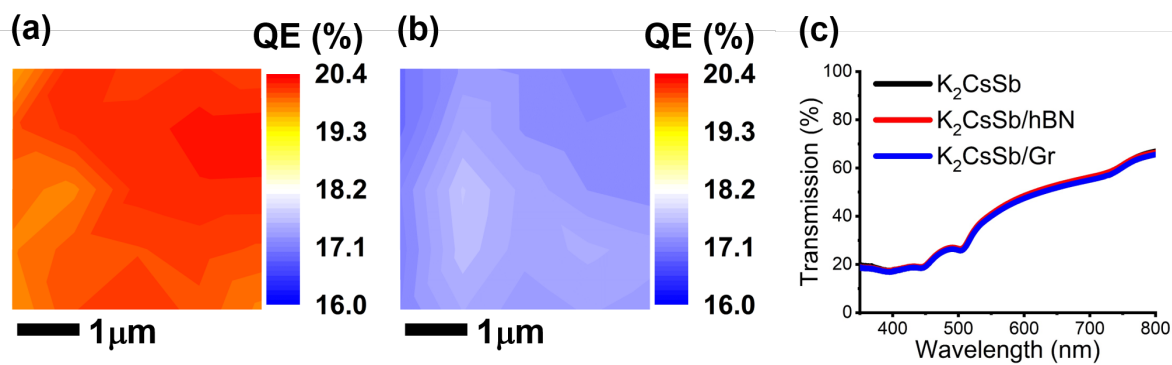

**Figure 4.** 405 nm illuminated QE map of $K_2CsSb$ photocathodes with (a) and without (b) hexagonal boron nitride monolayer coating on nickel and stainless steel substrates, respectively. Spatial resolution is 350 nm. Both $K_2CsSb$ photocathodes were deposited simultaneously. Scale bars are 1 μm. (c) Optical transmission spectra of $K_2CsSb$ photocathodes on bare, graphene (Gr) coated, and hexagonal boron nitride (hBN) coated sapphire substrates sealed in a vacuum tube.

Implications of these results are broad. A scientific implication is that there could be a novel optical interaction mechanism to enhance the QEs of semiconductor photocathodes using atomically thin coatings. This could open up a new pathway in the ongoing approach of engineering substrates to enhance the QE of deposited semiconductor photocathodes[13, 15]. A technological implication could be that atomically thin layer coating eliminate time-consuming optimizations of substrate preparations. Our results demonstrate that simply by coating atomically thin layers on metallic substrates, $K_2CsSb$ photocathodes with QEs higher than those on electrochemically polished and chemically passivated metal substrates can be achieved.

**2.2 Material characterization of photocathodes deposited on atomically thin 2D crystal coated substrates**

We previously demonstrated that $K_2CsSb$ photocathodes grown on graphene coated substrates exhibit X-ray diffraction (XRD) and X-ray fluorescence (XRF) spectra that are consistent with those obtained when using uncoated silicon substrates[31]. These results indicate that both the crystal quality and elemental stoichiometry of $K_2CsSb$ photocathodes do not appreciably change due to a graphene coating. Here, we performed a similar study using monolayer hBN films. At the National Synchrotron Light Source II (NSLS-II) of Brookhaven National Laboratory (BNL), we deposited $K_2CsSb$ photocathodes on hBN coated substrates and monitored XRD and XRF in-situ. The results are shown in **Figure 5** (a) and (b). Figure 5 (a) is the XRD spectra of $K_2CsSb$ photocathodes deposited with (red) and without (black) hBN films on sapphire substrates. Sapphire is known to be a reliable substrate material for $K_2CsSb$ photocathodes growth and we assumed that there should not be a crystal quality difference. Peak positions and their full width at half maximum (FWHM) are summarized in Table 1. The peak positions and FWHM were identical to each other with a good match of d-spacings to the theoretical values. Specifically, d-spacings were 3.40 and 2.15 Å, and FWHMs were 0.03 and 0.01 Å for (002) and (004) crystal orientations, respectively. These results indicate that the crystal quality of $K_2CsSb$ photocathodes does not change due to the hBN coating. This is consistent with elemental stoichiometry of close to $K_2CsSb$ achieved on hBN coated substrate.

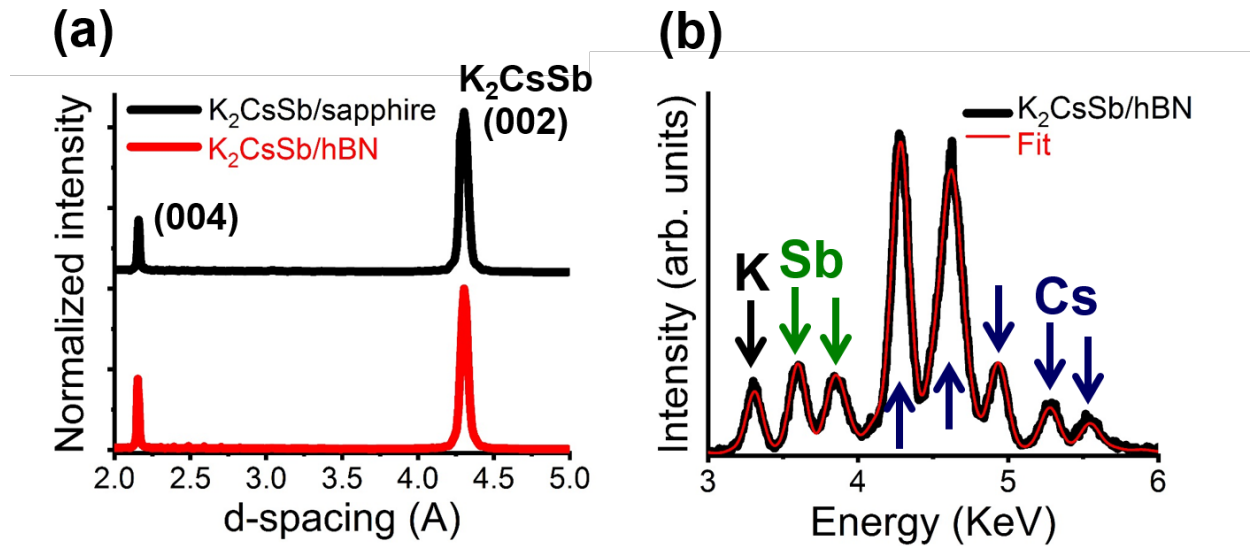

Figure 5(b) is XRF spectrum of the $K_2CsSb$ photocathodes grown on the hBN coating. Potassium (K), antinomy (Sb), and cesium (Cs) as present as expected, where the peak near 4.9 KeV may have minimal contribution from titanium (Ti) sample mount. Spectrum analysis revealed stoichiometry of $K_{2.37}Cs_{1.05}Sb$.

Based on such material characterizations of the photocathodes, the general appearance is that 2D crystal coatings do not alter their crystal quality and elemental stoichiometry compared to well-stablished substrates such as sapphire and silicon. It is likely that that the dangling bond-free atomic structure of 2D crystals provides a chemically inert surface for highly-crystalline photocathode growth.

**Figure 5.** (a) X-ray diffraction (XRD) spectra of $K_2CsSb$ photocathodes deposited on sapphire (black) and hexagonal boron nitride coated substrates (red) at the National Synchrotron Light Source II of Brookhaven National Laboratory. Intensity is normalized to the $K_2CsSb$ (002). (b) X-ray fluorescence (XRF) spectrum of hexagonal boron nitride coated substrate (black). Colored arrows indicate the peak positions of potassium (K), antinomy (Sb), and cesium (Cs),

|  | d-spacing (Å) | Width (Å) | (HKL) | Theory (Å) |
|---|---|---|---|---|
| **Cathode/sapphire** | 4.300 | 0.037 | (002) | 4.310 |
|  | 2.159 | 0.010 | (004) | 2.155 |
| **Cathode/hBN** | 4.305 | 0.033 | (002) | 4.310 |
|  | 2.155 | 0.010 | (004) | 2.155 |

respectively. Red line is the fitted spectrum used for quantitative analysis.

**Table 1.** List of d-spacing for observed peaks in comparison to the theoretical values for $K_2CsSb$. Corresponding crystal facet orientations and FWHM peak widths are also shown.

## 2.3 Wavelength dependence of quantum efficiency enhancement by atomically thin 2D crystal coating and a possible enhancement mechanism

*2.3.1 Wavelength dependence of quantum efficiency enhancement by atomically thin 2D crystal coating*

The wavelength dependence of the QE enhancement was investigated to gain insights into its mechanism. Specifically, lasers and LEDs with wavelengths of 375 nm, 405 nm, 532 nm, and 633 nm were focused to obtain QE maps of an identical region at the transition between bare and graphene coated stainless steel in the phototube. Spatial resolution of these QE maps is ~1 μm and are shown in **Figure 6**. The QE is normalized to highest values in each map, and the black arrows indicate the boundary of the graphene coated (top half) and non-coated regions (bottom half). It is instantly evident that the enhancement has a strong wavelength dependence and that the enhancement increases at longer wavelengths. While the enhancement is negligible at 375 nm, it becomes relative 10 % on average at 405 nm (consistent with the previous section), 18 % at 532 nm, and reaches 36 % at 633 nm. The maximum value in a "hot" spot in the 633 nm map is as high as an 80 % relative increase in QE.

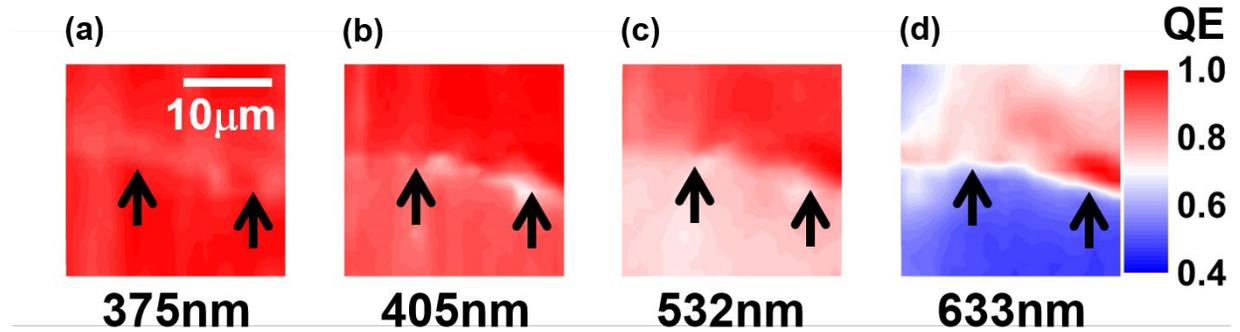

**Figure 6.** QE maps of $K_2CsSb$ photocathodes on an identical region where stainless steel substrate is half coated with graphene (top half). Four different illumination wavelengths of (a) 375 nm, (b) 405 nm, (c) 532 nm, and (d) 633 nm were used. Black arrows indicate the location of boundary between graphene coated and non-coated regions. Spatial resolution is ~1 μm and intensity are normalized to the maximum value. Scale bar is 10 μm and applies for all maps.

*2.3.2 Possible mechanisms of the quantum efficiency enhancement*

We considered the following possible mechanisms for the observed QE enhancement; 1) interference effects due to the 2D crystal coatings, and 2) enhancement of substrate mirroring effect due to physical gaps created between photocathodes and substrates by 2D crystal coatings. For 1), beneficial interference effects can occur when the particular optical constants and film thicknesses result in enhanced absorption and thus QE. This involves well-known thin film optical principles e.g. for design of light emitting diodes (LEDs) to enhance their brightness[32, 33], or antireflective coatings on lenses and windows. Total optical absorption by a photocathode due to interference effects is a strong factor of the reflectivity of substrate, thus it is possible that the 2D crystal coating changes the reflectivity of substrate such that QE of deposited photocathode increases. For 2), it is established knowledge that the electric field must decrease in the vicinity of a conducting surface to satisfy electromagnetic boundary conditions (and must vanish entirely for an ideally perfect conductor/reflector). Since the photocathode and 2D sublayer coating thicknesses are much smaller than optical wavelengths, the resultant absorption and QE for a photocathode in intimate contact with a highly reflective substrate is *lower* than that without the substrate, since the local electric field is reduced. Our investigations suggest that due to the substrate roughness, the 2D crystal coatings may actually create effective physical gaps of few hundred nanometers or more between the photocathode and metal substrate, and this may enhance the mirroring effect.

Our evaluation of the QE enhancement by mechanism 1) using a transfer matrix approach matched experimental results qualitatively but not quantitatively. Calculated values were roughly one order of magnitude lower than experimental values for wavelength range of 375-633 nm that we measured. We covered photocathode thickness of 15-25 nm, incident angle of 0-89 degrees as well as substrate materials of iron and silicon. All of these parameters have strong effects on optical absorption of photocathodes when conditions are in interference effects regime[13]. The fact that they did not alter the absorption by photocathodes in our case suggests that it is not the dominant mechanism to explain our results. It simply implies that the 2D crystal coatings do not modify the reflectivity of underlying substrates in our model, which is reasonable due to the relatively high matching of the optical constants.

On the contrary, we found that mechanism 2) could explain our results much better. Figure 7 (a), (b) are scanning electron microscope images of stainless steel substrate with and without graphene coatings, respectively. One can see graphene wrinkles that go horizontally across the image as indicated by white arrows. These suggests that graphene spans over the grooves in these mechanically ground or rolled substrates as schematically illustrated in Figure 7(c). Optical profilometer measurements of the substrates indicated average surface roughness of 230 nm, and grooves with average maximum height of 1.4 μm and pitch of 15-50 μm. We previously demonstrated that free-standing 2D crystals applied with these transfer techniques will naturally span large voids and still support $K_2CsSb$ photocathodes[6, 31] thus we reason it is most probable that there are effective physical gaps between photocathodes and the quite rough stainless steel substrates in this case as well.

Contrary to the case where a photocathode is in intimate and perfect contact with a reflecting surface, a physical gap between reflector and photocathode can result in absorption and QE increases. For ideal planar geometries one must consider coherent interference effects, but in the much more realistic case where surfaces are rough and incoherent, a simple multiple reflection enhancement is likely. Figure 7 (d) shows calculation results for the relative optical absorption enhancement (hence QE) for a $K_2CsSb$ photocathode with an ideal stainless steel reflecting mirror located underneath it, compared to that of a free-standing photocathode. Optical reflectivity of stainless steel in Zwinkels *et al.*[34] was used for the calculation. The model matches our experimental results qualitatively with an exception of 375 nm. The reflectivity of stainless steel decreases as wavelength gets shorter, especially around 375 nm[34] thus it might be responsible for the mismatch.

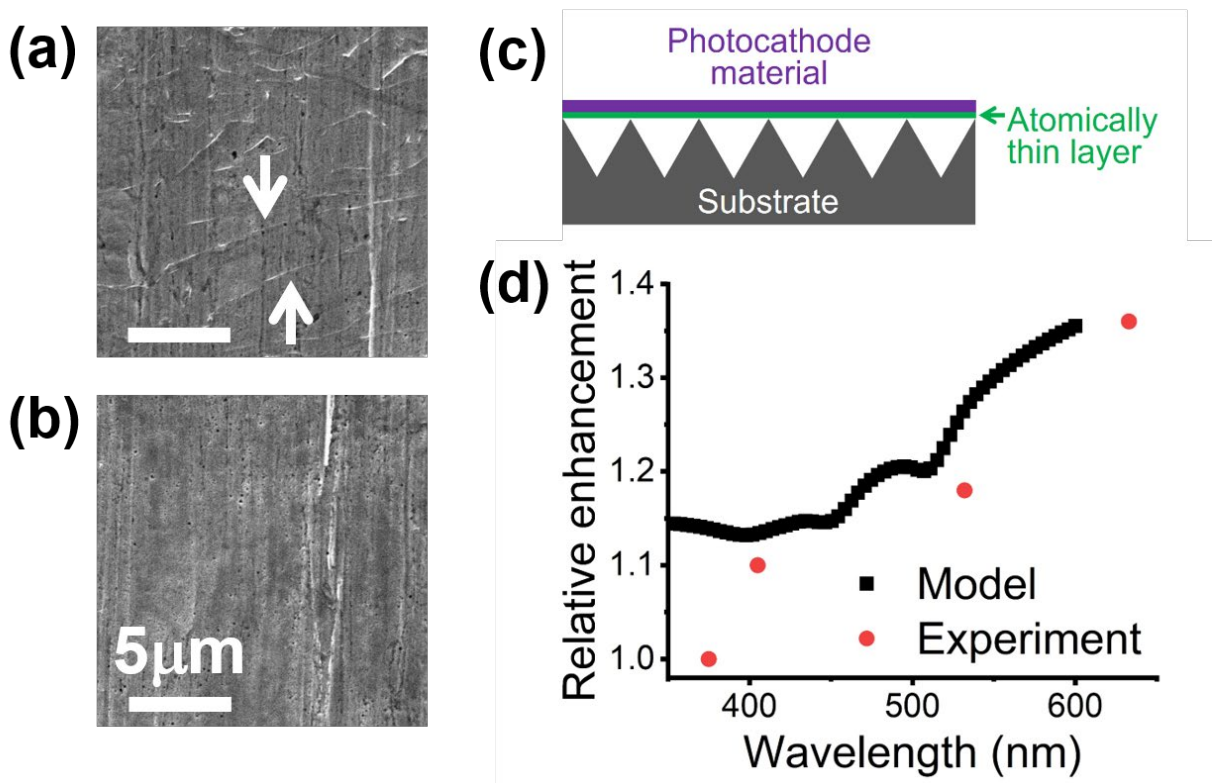

**Figure 7.** (a), (b) Scanning electron microscope (SEM) images of stainless steel substrates with and without graphene coating, respectively. Scale bar is 5 μm. White arrows indicate graphene wrinkles that go horizontally across the image. (c) Simplified illustration of Figure 7 (a). (d) Calculated relative optical absorption enhancement of photocathodes with a physical gap between photocathode and reflective substrate (black) compared with QE enhancement observed experimentally (red).

## 3. Conclusions

In summary, we observed unexpected QE enhancement of $K_2CsSb$ photocathodes by interfacing them with an atomically thin 2D crystal layer. The enhancement occurred in a reflection mode, when a 2D crystal was placed in between photocathodes and optically reflective substrates. Specific combinations of 2D crystal and reflective substrates that we investigated were graphene on stainless steel and hexagonal boron nitride monolayer on nickel. Observed QE enhancement was average of relative 10 % at ~3.1 eV (405 nm), relative 18.2 % at ~2.0 eV (532 nm), and relative 36 % at ~2.0 eV (633 nm) when the photocathodes were deposited on graphene coated stainless steel. Surprisingly, the enhancement reached up to relative 80 % at "hot" spot regions at ~2.0 eV. The enhancement was relative 9.0 % on average at ~3.1 eV for photocathode on hexagonal boron nitride coated nickel. The enhancement did not occur when reflective substrates were replaced with optically transparent sapphire. Optical transmission spectra of $K_2CsSb$ photocathodes on bare, graphene coated, and hexagonal boron nitride coated sapphire substrates were identical, which indicates that $K_2CsSb$ photocathode thickness does not change due to 2D crystal coatings. XRD and XRF revealed that crystal orientation, quality and elemental stoichiometry do not appreciably change either due to 2D crystal coatings. Surface morphology of 2D crystal coated substrates indicated that 2D crystals span over the grooves of grinded substrates, which suggests that 2D crystal coatings create effective physical gap between deposited photocathodes and substrate surfaces. Our model indicates that this physical gap enhances the mirroring effect of reflective substrates thus leads to QE enhancement we observed. The results provide a pathway toward simple method to enhance the QE of semiconductor photocathodes by an atomically thin 2D crystal on substrates.

# Experimental Section

***Synthesis and transfer of atomically thin 2D crystals:*** Graphene monolayers were synthesized *via* chemical vapor deposition (CVD) using methane gas as the carbon source and copper (Cu) foils as substrates. CVD monolayer hexagonal boron nitride (hBN) grown on copper foil was purchased from Graphene Supermarket and was incorporated into the phototubes. Monolayer hexagonal boron nitride used for the photocathode material characterization was synthesized *via* the epitaxial growth on sapphire substrates by low-pressure CVD[35]. For wet-transfer of CVD graphene and hexagonal boron nitride monolayer onto various substrates, poly(methyl methacrylate) (PMMA) was used as a mechanical support and removed by subsequent acetone rinsing.

***Deposition of bialkali antimonide photocathodes for vacuum tubes:*** Graphene films on a stainless steel foil frame (SS304), hexagonal boron nitride films on nickel (Ni) TEM mesh grid, and both films on annealed sapphire were installed in phototube assemblies for bialkali antimonide photocathode deposition at Photonis Defense Inc. All materials in the vacuum envelope were pre-cleaned in-situ at 350 °C in UHV prior to photocathode deposition. For the annealed sapphire with graphene and hexagonal boron nitride coatings, ex-situ annealing at 600 °C in hydrogen gas atmosphere were also performed to chemically reduce the sealing metal. While monitoring the sensitivity of the photocathode films, the components K, Cs, and Sb were deposited on substrates *via* thermal evaporation to achieve typical stoichiometry of $K_2CsSb$ with thickness of ~20 nm. The vacuum-sealed packages consisted of sapphire windows on both sides of the photocathode assembly with metal traces patterned on the windows to establish an extracting electric field.

***Photoemission measurement of bialkali antimonide photocathodes in vacuum tubes:*** 375 nm pulsed laser, 405 and 532 nm light emitting diodes (LEDs), 633 nm He-Ne laser, and Fianium WhiteLase tunable laser (400-2,400 nm, repetition rate 40 MHz) equipped with a Fianium AOTF (Acousto-Optic Tunable Filter) system were used as light sources for photoemission measurements. The focused spot size for large area QE maps was ~0.20 mm (405 nm LED), <5 μm for wavelength dependence QE maps (375 nm pulsed laser, 405 and 532 nm LEDs, and 633 nm He-Ne laser), and < 350 nm for a high spatial resolution QE maps (Fianium WhiteLase tunable laser). Spatial resolution for the wavelength dependence QE maps was ~1 μm as we used 1 μm steps. Anode traces on the sapphire windows were sufficiently biased with respect to the photocathode assembly to overcome space-charge effects and collect photoelectrons in all cases. The quantum efficiency was calculated using the known power of incident light at each wavelength, as obtained from a calibrated reference diode. A home-built confocal microscopy system with a scanning mirror that allows for precise location of the focal point onto the sample surface was used for high spatial resolution QE maps.

*Material characterization of bialkali antimonide photocathodes:* In-situ X-ray diffraction (XRD) growth studies on $K_2CsSb$ were performed at the Brookhaven National Laboratory National Synchrotron Light Source II (NSLS-II) beamline ID-4 (ISR) using photon energy of 11.47 KeV ($\lambda$ = 1.0809 Å). The thin film growth was performed in a custom-built ultrahigh vacuum chamber with a base pressure of low $10^{-10}$ Torr. Hexagonal boron nitride monolayers grown by CVD were transferred onto silicon (Si) substrates. The reference Si substrates and coated substrates were loaded into the growth chamber and annealed at 550 ºC for 1 hour. Co-evaporation of K, Cs and Sb using pure metallic sources was used to fabricate $K_2CsSb$ photocathodes. The evaporation rate was controlled by adjusting the current of the fusion cells and was measured with a quartz crystal microbalance (QCM) placed alongside the sample. Alkali and antimony sources were turned on simultaneously, and the rates of the three were set to match the stoichiometry of $K_2CsSb$ based on real-time X-ray fluorescence (XRF) analysis. During deposition, the substrate temperature was set to about 90 ºC. The XRD data were measured using a 4 axis diffractometer with a Pilatus 100 K X-ray camera mounted 70 cm downstream from the substrate. XRD was measured with a 2θ range from 5° to 25°. The XRF spectra were measured by a vortex multi-cathode X-ray detector mounted 45° with respect to the sample surface normal and approximately 25 cm away from the sample.

*Optical transmission measurements of bialkali antimonide photocathodes in a vacuum tube:* Jasco V-730 UV-Visible Spectrophotometer with light spot of 1x2 mm was used to measure optical transmission spectra of bialkali antimonide photocathodes that are on bare and atomically thin 2D crystal coated sapphire substrates.

***Absorption calculation of bialkali antimonide photocathodes:*** Calculations for interference effects analysis were performed using a transfer matrix approach assuming normal incidence[36]. Optical constants of graphene and Si were taken from Kravets[37] and Vuye[38], respectively. Optical constants of Fe from Johnson and Cristy were used to estimate stainless steel[39]. Optical constants for $K_2CsSb$ were provided by Photonis and were consistent with calculations using optical constants from Motta[40]. Calculations for substrate mirroring effect analysis were performed using NKD Gen software (available from University of Barcelona), which is also based on conventional transfer matrix methodology. Established literature values were used for optical constants (n,k) of $K_2CsSb$ and stainless steel[34].

## Acknowledgments

The work was financially supported by the U.S. Department of Energy (DOE) Office of Science U.S.-Japan Science and Technology Cooperation Program in High Energy Physics. This research used the ISR beamline of the National Synchrotron Light Source II, a U.S. Department of Energy (DOE) Office of Science User Facility operated for the DOE Office of Science by Brookhaven National Laboratory under Contract No. DE-SC0012704. In particular, the authors wish to thank Ken Evans-Lutterodt and John Walsh for their expert assistance.

## Conflict of Interest

H.Yamaguchi, F.Liu and N.A. Moody have submitted provisional patent on the topic of this study. The other authors declare that they have no financial competing interests.